\documentclass[pre,preprint,showpacs,preprintnumbers,amsmath,amssymb]{revtex4}
\usepackage{graphicx}
\usepackage{latexsym}
\usepackage{bm}
\usepackage{epsfig}
\usepackage{amssymb}

\begin{document}

\title{Callen--like method for the classical Heisenberg ferromagnet}

\author{L. S. Campana}
\author{U. Esposito}
\affiliation{Dipartimento di Scienze Fisiche, Universit\`a degli Studi di Napoli ``Federico II'', Piazzale Tecchio 80, I-80125 Napoli, Italy}

\author{ A.~Cavallo} 
\affiliation{Dipartimento di Fisica "E.R. Caianiello", Universit\`a degli Studi di Salerno, via Ponte don Melillo, I-84084 Fisciano (Salerno), Italy}

\author{L. De Cesare}
\author{A. Naddeo}
\affiliation{Dipartimento di Fisica "E.R. Caianiello", Universit\`a degli Studi di Salerno, via Ponte don Melillo, I-84084 Fisciano (Salerno), Italy\\
CNISM, Unit\`a di Salerno, I-84084 Fisciano (Salerno), Italy}

\date{\today}

\begin{abstract}
A study of the $d$-dimensional classical Heisenberg ferromagnetic model in the presence of a magnetic field is performed within the two-time Green function's  framework in classical statistical physics. We extend the well known quantum Callen method to derive analytically a new formula for magnetization. Although this formula is valid for any dimensionality, we focus on one- and three- dimensional models and compare the predictions with those arising from a different expression suggested many years ago in the context of the classical spectral density method. Both frameworks give results in good agreement with the exact numerical transfer-matrix data for the one-dimensional case and with the exact high-temperature-series results for the three-dimensional one. 
In particular, for the ferromagnetic chain, the zero-field susceptibility results are found to be consistent with the exact analytical ones obtained by M.E. Fisher. 
However, the formula derived in the present paper provides more accurate predictions in a wide range of temperatures of experimental and numerical interest.
	\end{abstract}
\pacs{ 75.10.Hk, 05.20.-y, 75.40.Cx}

\maketitle
	\section{Introduction}
	Classical spin models are widely studied in statistical mechanics and play an important role in condensed matter physics but also in other disciplines, such as biology, neural networks, etc.
		Despite their intrinsic simplicity with respect to the quantum counterpart, classical spin models show highly nontrivial features as, for instance, a rich phase diagram and finite-temperature criticality.

		Remarkably, in many relevant situations, the investigation of classical spin models allows to obtain several information for realistic magnetic materials. Indeed, they have turned out to be extremely versatile for describing a variety of relevant phenomena.
This justifies the  significant effort which has been devoted for optimized implementations of Monte Carlo simulations of spin models. Besides, there are several questions related to classical spin models which may further stimulate in developing new available computational resources, more efficient algorithms and powerful techniques for obtaining satisfactory answers.

		Along this direction, several methods have been employed to investigate different classical spin models such as Ising, Potts models and several variants of the basic Heisenberg model.

		A very efficient tool of investigation, in strict analogy of the quantum many-body techniques, is constituted by the two-time Green functions (GF) framework in classical statistical physics, as suggested by Bogoliubov Jr. and Sadovnikov \cite{bogoliubov} some decades ago and further developed and tested in Refs. \cite{prof1984,cavallo,prof1,rassegna2006}.

The central problem in applying this method to quantum~\cite{quantumspinmodels} and classical~\cite{prof1984,cavallo,prof1, rassegna2006,prb2010} spin models is to find a suitable expression for magnetization in terms of the basic two-time Green function for the specific spin Hamiltonian under study.

The same problem occurs in the related quantum~\cite{quantumSDM} and classical~\cite{bogoliubov,prof1984,cavallo,prof1,rassegna2006} spectral density methods (QSDM and CSDM) which have been less  used in literature although they appear very effective to describe the macroscopic properties of a variety of many body systems~\cite{rassegna2006}. 

For quantum spin-$1/2$ Heisenberg models, exact spin operator relations allow to solve directly the problem. The case of spin-S was solved in an elegant way by Callen~\cite{callen1963} (in the context of the two-time GF method) providing a general formula for magnetization, successfully used in many theoretical studies.

Unfortunately, for classical Heisenberg models, a similar formula is lacking in the classical two-time GF framework~\cite{bogoliubov,prof1984,cavallo,prof1,rassegna2006}. This difficulty  is related to the absence of a kinematic sum rule for the $z$-component of the spin vector as it happens in quantum counterpart.

Almost three decades ago~\cite{prof1984}, in the context of the CSDM~\cite{prof1984,cavallo,prof1,rassegna2006}, a suitable formula for magnetization was suggested  for the classical isotropic Heisenberg model. 
However, its analytical structure was conjectured on the ground of known asymptotic results for near-polarized and paramagnetic states and not derived by means of general physical arguments.
In spite of this, also in the intermediate regimes of temperature and magnetic field, this formula provided results in very good agreement with the exact numerical transfer matrix (TM)~\cite{tognetti1,tognetti2} ones for a spin chain and with the exact high-temperature-series (HTS) data of Rushbrooke et al.~\cite{rush1}, for the three-dimensional model.

Quite recently~\cite{prb2010}, a study on a class of spin models based on the classical Heisenberg Hamiltonian has provided clear evidence of the effectiveness of the old formula for magnetization suggested in Ref.~\cite{prof1984} and of the CDSM itself to describe magnetic properties in a wide range of temperatures, in surprising agreement with high resolution simulation predictions. On this grounds, the authors were also able to explain some puzzling experimental features, confirming again that this formula appears to work surprisingly well in different contexts.

In this paper we are primarily concerned with this relevant question by accounting for an extension of the quantum Callen method to derive, in the context of the classical two-time GF-framework, and, on the ground of general arguments, an expression for magnetization of the $d$-dimensional classical Heisenberg model, which is just the classical analogous of the famous quantum Callen formula. Remarkably, the formula here derived reproduces the asymptotic results obtained within the CSDM~\cite{prof1984} in the near-polarized and near-zero magnetization regimes.

To test the most reliability of our formula, here we limit ourselves to calculate relevant quantities, as the spontaneous magnetization and critical temperature for $d=3$ and other ones for $d=1$, and to compare our predictions with those obtained in Ref.~\cite{prof1984}. Noteworthy is that, very small deviations are found in the intermediate regime corroborating the accuracy of the formula for magnetization suggested many years ago~\cite{prof1984} and, in turn, the robustness of that found here.

The paper is organized as follows. First, in Sec.~\ref{model}, we introduce the classical spin model and the appropriate classical Callen-like two-time Green function  with the related equation of motion which are the  main ingredients for next developments. Besides, we introduce, in a unified way, the Tyablikov- and Callen-like decouplings for higher order Green functions. In Sec.~\ref{moments} we extend the quantum Callen approach to derive the general formula for magnetization valid for arbitrary dimensionality, temperature and applied magnetic field. It is easily obtained overcoming the intrinsic difficulty related to the feature that the classical analogous of the quantum spin identities used by Callen~\cite{callen1963} does not exist. In Sec.~\ref{equations}, self-consistent equations are obtained allowing to determine the magnetization and hence other thermodynamic quantities of experimental and numerical interest. Calculations for magnetization, transverse correlation length and critical temperature for $d>2$ and different lattice structure are presented in Sec.~\ref{calculations}. Finally, in Sec.~\ref{conclusions}, some conclusions are drawn.

\section{The model and Callen-like Green function}~\label{model}
The classical Heisenberg ferromagnet is described by the Hamiltonian
\begin{eqnarray}
\cal{H}&=&-\frac{1}{2}\sum_{i,j=1}^{N}J_{ij}{\bf S}_i\cdot{\bf S}_j-h\sum_{i=1}^{N}S_i^z\nonumber\\
&=&-\frac{1}{2}\sum_{i,j=1}^{N}J_{ij}\left(S_i^+S_j^-+S_i^zS_j^z\right)-h\sum_{i=1}^{N}S_i^z.\label{eq1}
\end{eqnarray}
Here $N$ is the number of sites on a $d$-dimensional lattice with unitary spacing, $\left\{{\bf S}_j\equiv\left(S_j^x,S_j^y,S_j^z\right);i=1,2,...,N\right\}$ are classical spin vectors with $\left|{\bf S}_j\right|=S$, $S_j^\pm=S_j^x\pm i S_j^y$, $J_{ij}=J_{ji}$, $(J_{ii}=0)$ is the spin-spin coupling and $h$ is the applied magnetic field. Of course, the identity ${\bf S}_j^2=\left(S_j^z\right)^2+S_j^+S_j^-=S^2$ is valid. For formal simplicity, in the next developments we will assume $S=1$. This assumption is perfectly legal in the classical context.

The model~(\ref{eq1}) can be appropriately described in terms of the $2N$ canonical variables $\phi\equiv\left\{\phi_j\right\}$ and $S^z\equiv\left\{S_j^z\right\}$, where $\phi_j$ is the angle between the projection of the spin vector ${\bf S}_j$ in the $x-y$-plane and the $x$-axis.

The basic spin Poisson brackets are
\begin{eqnarray}
\left\{S_i^z,S_j^{\pm}\right\}&=&\mp i S_i^{\pm}\delta_{ij},\nonumber\\
\left\{S_i^+,S_j^-\right\}&=&-2 i S_i^z\delta_{ij},\label{eqpoissonbr}\\
\left\{S_i^\alpha,S_j^{\beta}\right\}&=&\epsilon_\gamma^{\alpha\beta} S_i^{\gamma}\delta_{ij},\quad (\alpha,\beta,\gamma=x,y,z),\nonumber
\end{eqnarray}
where $\left\{...,...\right\}$ denotes a Poisson bracket and $\epsilon_\gamma^{\alpha\beta}$ is the Levi-Civita tensor.

In strict analogy with the quantum counterpart~\cite{callen1963}, we now introduce the classical two-time retarded GF~\cite{bogoliubov,rassegna2006}
\begin{eqnarray}
G_{ij}(t-t')&=&\theta(t-t')\langle\left\{S_i^{+}(t-t'),e^{aS_j^z}S_j^-\right\}\rangle \nonumber\\
&=& \langle\langle S_i^{+}(t-t');e^{aS_j^z}S_j^-\rangle\rangle.\label{eqgf}
\end{eqnarray}
Here $\theta(x)$ is the usual step function, $a$ is the Callen-like parameter, $\langle ...\rangle=Z^{-1} \prod_j\int_0^{2\pi}d\phi_j\int_{-1}^{1} ...d S_j^z e^{-\beta \mathcal{H}\left(\left\{\phi_j\right\},\left\{S_j^z\right\}\right)}$ stands for the usual statistical average, $\left\{  A,B\right\}$ denotes the Poisson bracket of the dynamical variables $A(\phi,S^z)$ and $B(\phi,S^z)$, and $X(t)=e^{iLt}X$ where $L...=i\left\{\mathcal{H},...\right\}$ is the Liouville operator. Of course $e^{iLt}$ acts as a classical time-evolution operator which transforms the dynamical variable $X\equiv X(0)\equiv X(\phi(0),S^z(0))$ at the initial time $t=0$ into the dynamical variable $X(t)\equiv X(\phi(t),S^{z}(t))$ at the time $t$.

The equation of motion (EM) for the GF (\ref{eqgf}) is given by (with $\tau=t-t'$)~\cite{rassegna2006}:
\begin{equation}
	\frac{dG_{ij}(\tau)}{d\tau} =
        \delta(\tau)\langle\left\{S_i^+,e^{a S_j^z} S_j^-\right\}\rangle+
        \langle\langle\left\{S_i^+(\tau),\mathcal{H}\right\};e^{aS_j^z} S_j^-\rangle\rangle,\label{eqem}
	\end{equation}

which, in the frequency-$\omega$ Fourier space, becomes:
\begin{equation}
\omega G_{ij}(\omega)=i\langle\left\{S_i^+,e^{aS_j^z} S_j^-\right\}\rangle+i\langle\langle\left\{S_i^+(\tau),\mathcal{H}\right\};e^{aS_j^z} S_j^-\rangle\rangle_\omega\label{eqemfourier}
 \end{equation}
with $G_{ij}(\omega)=\langle\langle S_i^+(\tau);e^{aS_j^z} S_j^-\rangle\rangle_\omega$ and $\langle\langle A(\tau);B\rangle\rangle_\omega = \int_{-\infty}^{+\infty}{ d}\tau e^{i\omega \tau}\langle\langle A(\tau);B\rangle\rangle$. 

Now, from basic Poisson brackets (\ref{eqpoissonbr}),
employing a straightforward algebra yields:
\begin{equation}
i\langle\left\{S_i^+,e^{aS_j^z} S_j^-\right\}\rangle=\psi(a)\delta_{ij},
\end{equation}
where
\begin{equation}
\psi(a)=i\langle\left\{S_i^+,e^{aS_i^z} S_i^-\right\}\rangle=-a\Omega(a)+2\Omega^{\prime}(a)+a\Omega^{\prime\prime}(a),\label{eqpsi}
\end{equation}
and
\begin{equation}
\Omega(a)=\langle e^{aS_i^z}\rangle.
\end{equation}
On the other hand, in Eq.~(\ref{eqem}), we easily have also:
\begin{equation}
\left\{S_i^+,\mathcal{H}\right\}=i\sum_{h}J_{ih}\left[S_i^z S_h^+ - S_i^+ S_h^z\right]-ihS_i^+.
\end{equation}
Then Eq.~(\ref{eqemfourier}) becomes
\begin{eqnarray}
& (\omega-h)G_{ij}(\omega) = \psi(a)\delta_{ij}+\nonumber\\
& -\sum_{h}J_{ih}\left[\langle\langle S_i^z(\tau) S_h^+(\tau);e^{aS_j^z} S_j^-\rangle\rangle_{\omega} - \langle\langle S_i^+(\tau) S_h^z(\tau);e^{aS_j^z} S_j^-\rangle\rangle_\omega\right]\label{eqemfourier2}.
 \end{eqnarray}
It is worth noting that, for magnetization per spin $\sigma=\langle S_i^z\rangle$ we have:
\begin{equation}
\sigma=\frac{1}{2}\psi(0).
\end{equation}
Up to this stage, no approximations are involved. 

For our purposes, to close Eq.~(\ref{eqemfourier2}) we consider here the following decouplings for the higher order-GF's:
\begin{equation}
\langle\langle S _h^z(\tau)S_k^+(\tau);e^{aS_j^z} S_j^-\rangle\rangle_\omega\approx \sigma\left[G_{kj}(\omega)-\lambda\frac{\langle S_{h}^{-}S_{k}^+\rangle}{2}G_{hj}(\omega)\right],
\end{equation}
where $\lambda=0$ (TD) and $\lambda=1$ (CD) correspond to the Tyablikov- and Callen-like decouplings. At this level of approximation, Eq.~(\ref{eqemfourier2}) reduces to the closed equation:
\begin{eqnarray}
& (\omega-h)G_{ij}(\omega)=\psi(a)\delta_{ij}-\sigma\sum_h J_{ih}\left[G_{hj}(\omega)-G_{ij}(\omega) \right]+\nonumber\\
& +\lambda\frac{\sigma}{2}\sum_h J_{ih}\left[\langle S_i^- S_h^+\rangle G_{ij}(\omega)-\langle S_h^- S_i^+\rangle G_{hj}(\omega)\right].\label{eqemfourier3}
 \end{eqnarray}
We now define~\cite{callen1963}, the ${\bf k}$-wave vector Fourier transforms in the first Brillouin zone as:
\begin{eqnarray}
G_{ij}(\omega)&=&\frac{1}{N}\sum_{{\bf k}}e^{i{\bf k}\cdot ({\bf r}_{i}-{\bf r}_{j})}G_{\bf k}(\omega),\nonumber\\
J_{ij}&=&\frac{1}{N}\sum_{{\bf k}}e^{i{\bf k}\cdot ({\bf r}_{i}-{\bf r}_{j})} J({\bf k}),\label{eqfouriertr}\\
C_{ij}(\omega)&=&\langle S_i^+ S_{i}^-\rangle =\frac{1}{N}\sum_{{\bf k}}e^{i{\bf k}\cdot ({\bf r}_{i}-{\bf r}_{j})}C({\bf k}).\nonumber
\end{eqnarray}
By replacing~(\ref{eqfouriertr}) in~(\ref{eqemfourier3}), in the $({\bf k},\omega)$-Fourier space the resulting equation for $G_{\bf k}(\omega)$ can be immediately solved and we find:
\begin{equation}
G_{\bf k}(\omega)=\frac{\psi(a)}{\omega-\omega_{\bf k}}, \label{eqGw}
\end{equation}
where
\begin{equation}
\omega_{\bf k}=h+\sigma(J({\bf 0})-J({\bf k}))+\lambda\frac{\sigma}{2}\frac{1}{N}\sum_{{\bf k}^{\prime}}\left[J({\bf k}^{\prime})-J({\bf k}-{\bf k}^{\prime})\right]C({\bf k}^{\prime}),\label{eqwk}
\end{equation}
defines the dispersion relation for the undamped classical oscillations in the systems in terms of the Fourier transform $J({\bf k})$ of the spin exchange coupling $J_{ij}$.

At this stage, in order to determine $G_{\bf k}(\omega)$, and hence the relevant
magnetic properties of the classical HM as functions of $T$ and $h$ at $a=0$ (as for instance $\sigma=\psi(0)/2$), one must obtain an explicit expression of $\psi(a)$ in terms of the basic correlation functions related to the original GF. A direct calculation of $\sigma$ constitutes an intrinsic difficulty for classical spin models since the classical counterpart of the quantum kinematic rule for the $z$ component of the spin does not exist.

Here, we extend to the classical HM the well known Callen method for calculation of $\psi(a)$ and hence of magnetization which is the key of the next thermodynamic analysis. This will be the main subject of the next section.

	\section{Callen-like approach for magnetization}~\label{moments}

We first note that, from Eq.(\ref{eqGw}) and the known relation for the spectral density $\Lambda_{\bf k}(\omega)$, corresponding to the Green function $G_{\bf k}(\omega)$~\cite{rassegna2006},
\begin{equation}
\Lambda_{\bf k}(\omega)=i\left[G_{\bf k}(\omega+i\epsilon)-G_{\bf k}(\omega-i\epsilon)\right],\quad\epsilon\rightarrow 0^{+},
\end{equation}
we find
\begin{equation}
\Lambda_{\bf k}(\omega)=2\pi\psi(a)\delta(\omega-\omega_{\bf k}).
\end{equation}
Then, it is easy to derive the spectral density $\Lambda_{ij}(\omega)$ for $G_{ij}(\omega)$ by using the Fourier representation
\begin{eqnarray}
\Lambda_{ij}(\omega)&=&\frac{1}{N}\sum_{{\bf k}}e^{i{\bf k}\cdot ({\bf r}_{i}-{\bf r}_{j})}\Lambda_{\bf k}(\omega)\nonumber\\
&=& 2\pi \psi(a) \frac{1}{N}\sum_{{\bf k}}e^{i{\bf k}\cdot ({\bf r}_{i}-{\bf r}_{j})}\delta(\omega-\omega_{\bf k}).
\end{eqnarray}
Now, we can calculate the correlation function $\langle e^{aS_j^z} S_j^- S_i^+\rangle$ associated to the initial GF $G_{ij}(\omega)=\langle\langle S_i^+(\tau);e^{aS_j^z} S_j^-\rangle\rangle_\omega$~\cite{rassegna2006}. 

From the classical spectral theorem~\cite{rassegna2006}
\begin{equation}
\langle BA\rangle=T\int_{-\infty}^{+\infty}\frac{ d\omega}{2\pi}\frac{\Lambda_{AB}(\omega)}{\omega},
\end{equation}
we get
\begin{eqnarray}
\langle e^{aS_j^z}S_j^- S_i^+\rangle&=&T\int_{-\infty}^{+\infty}\frac{d\omega}{2\pi}\frac{\Lambda_{ij}(\omega)}{\omega}\nonumber\\
&=& T\psi(a)\frac{1}{N}\sum_{{\bf k}}\frac{e^{i{\bf k}\cdot ({\bf r}_{i}-{\bf r}_{j})}}{\omega_{\bf k}},
\end{eqnarray}
which implies also that, in Eq. (\ref{eqwk}),
\begin{equation}
C({\bf k})=\langle S_{\bf k}^{+} S_{-\bf k}^{-}\rangle=\frac{T\psi(0)}{\omega_{{\bf k}}}=T\frac{2\sigma}{\omega_{{\bf k}}}.\label{eqck}
\end{equation}
In particular, for $i=j$, we have
\begin{equation}
\langle e^{aS_i^z}S_i^- S_i^+\rangle=\Phi\psi(a),\label{eqckmod}
\end{equation}
where the quantity
\begin{equation}
\Phi=\frac{T}{N}\sum_{{\bf k}}\frac{1}{\omega_{\bf k}}\stackrel{N\rightarrow \infty }{=}T\int_{1BZ}\frac{d^{d}k}{(2\pi)^{d}}\frac{1}{\omega_{\bf k}},\label{eqphi}
\end{equation}
does not depend on the Callen-like parameter $a$. On the other hand, by using the identity $S_i^{+}S_i^{-}=S_i^{-}S_i^{+}=1-(S_i^{z})^{2}$, the correlation function on the left side of Eq.~(\ref{eqckmod})
can be independently expressed in terms of the function $\Omega(a)=\langle e^{aS_i^z}\rangle$ as
\begin{equation}
\langle e^{aS_i^z}S_i^- S_i^+\rangle=\Omega(a)-\Omega^{\prime\prime}(a).
\end{equation}
Then, with Eq.~(\ref{eqpsi}), we finally obtain for $\Omega(a)$ the equation
\begin{equation}
\Omega^{\prime\prime}(a)+2\left(\frac{1}{\Phi}+a\right)^{-1}\Omega^{\prime}(a)-\Omega(a)=0,\label{eqomegadiff}
\end{equation}
for which the initial condition $\Omega(0)=1$ is valid by definition. This is, of course, insufficient to find the physical solution of Eq.~(\ref{eqomegadiff}) and one should add a supplementary condition to be searched properly. Unfortunately, there is no classical analogue of the operatorial identity $\Pi_{p=-S}^{S}\left(S^{z}-p\right)=0$ which is the key ingredient of the Callen approach for the spin-$S$ quantum HM~\cite{callen1963}. In the following, we will show that, at our level of approximation, the additional  condition
\begin{equation}
\Omega(a)=\int_{-1}^{1}dS^{z}f(S^{z})e^{aS^{z}},\label{eqomegastatav}
\end{equation}
which follows formally from the definition of the canonical ensemble average of the dynamical variable $e^{aS^{z}}$, combined with $\Omega(0)=1$, allows to determine univocally $\Omega(a)$.

In view of the structure of the differential equation~(\ref{eqomegadiff}) for $\Omega(a)$, we assume
\begin{equation}
f(S^{z})=g(S^{z})e^{S^{z}/\Phi},\label{eqfSz}
\end{equation}
so that, we can write
\begin{equation}
\Omega(a)= \int_{-1}^{1}dS^{z} \; g(S^{z})e^{\left(\frac{1}{\Phi}+a\right)S^{z}}.\label{eqomegaint}
\end{equation}
It is worth noting that, in terms of $y = 1/\Phi + a$ $\left(\Omega(a) \Longleftrightarrow \Omega(y)\right)$,  by introducing $\rho(y)= y\Omega(y)$  Eq.~(\ref{eqomegadiff}) reduces to $\rho^{\prime\prime}(y) - \rho(y) =0$ which can be simply solved by means of a combination of hyperbolic sine and cosine functions. So, one easily obtains the general solution $\Omega(y) = A \sinh(y)/y + B \cosh(y)/y$ where the constants $A$ and $B$ have to be determined by using the initial condition $\Omega(1/\Phi) = 1$ at $a=0$ and the average-value nature of $\Omega(y)$, as expressed by the additional condition ~(\ref{eqomegastatav}). 
Indeed, it is immediate to see that the odd solution $\cosh(y)/y$ is incompatible with this last requirement (it should require a non-positive weighting function $g(S^z)$ introduced in Eq.~(\ref{eqomegaint})), so that  it must necessarily be  $B=0$.  Besides, from $\Omega(1/\Phi) = 1$ one easily obtains $A=(1/\Phi)/\sinh(1/\Phi)$ and 
and hence, as a function of the Callen-like parameter, the physical solution of the original Eq.(26) takes the form:

\begin{equation}
\Omega(a) = \frac{1/\Phi}{1/\Phi + a}\frac{\sinh\left[\frac{1}{\Phi}+a\right]}{\sinh\left(\frac{1}{\Phi}\right)}.
\end{equation}
This is the central result of the paper which constitutes the classical analogue of the quantum Callen formula~\cite{callen1963}. It provides for the magnetization $\sigma$ the expression (which is valid for any $d$, $T$ and $h$)
\begin{equation}
\sigma=\left[\coth\left(\frac{1}{\Phi}\right)-\Phi\right],\label{eqm}
\end{equation}
where $L(x)=\coth x -\frac{1}{x}$ is the Langevin function. Here $\Phi$ is given by Eq.~(\ref{eqphi})
in terms of the classical oscillation spectrum $\omega_{\bf k}$ which depends on $\sigma$ itself. Hence, Eq.~(\ref{eqm}) is a self-consistent equation for $\sigma$  where $T$ and $h$ enter the problem through $\Phi$.

\section{Self-consistent equations}\label{equations}

For next developments, it is convenient to work in terms of the
dimensionless quantities
\begin{equation}
\begin{array}{ccccc}
\gamma _{\bf{k}}=\frac{J\left( \bf{k}\right) }{%
J\left( 0\right) }, & \widetilde{\omega }_{\bf{k}}=\frac{\omega
_{\bf{k}}}{J\left( 0\right)}, & \widetilde{T}=\frac{T}{J\left(
0\right)}, & \widetilde{h}=\frac{h}{J\left( 0\right)}.
\end{array}
\label{sc1}
\end{equation}
Then, the basic equations to determine the spectrum of elementary
excitations and the magnetization become
\begin{equation}
\left\{
\begin{array}{c}
\widetilde{\omega }_{\bf{k}}=\widetilde{h}+\sigma \left(
1-\gamma _{\bf{k}}\right) R\left( \bf{k}\right)  \\
\sigma =L\left( \frac{1}{\Phi}\right) =\coth\left( \frac{1}{%
\Phi}\right) -\Phi
\end{array}
\right. ,  \label{sc2}
\end{equation}
with
\begin{equation}
R\left( \bf{k}\right) =1+\lambda \sigma \widetilde{T}\int_{1BZ}%
\frac{d^{d}k^{^{\prime }}}{\left( 2\pi \right) ^{d}}\frac{\gamma _{%
\bf{k}^{^{\prime }}}-\gamma _{\bf{k}-\bf{%
k}^{^{\prime }}}}{\left( 1-\gamma _{\bf{k}}\right) \widetilde{%
\omega }_{\bf{k}^{^{\prime }}}},  \label{sc3}
\end{equation}
and
\begin{equation}
\Phi =\widetilde{T}\int_{1BZ}\frac{d^{d}k}{\left(
2\pi \right) ^{d}}\frac{1}{\widetilde{\omega }_{\bf{k}}}.
\label{sc4}
\end{equation}
Although previous equations are true for generic 
center-symmetric exchange interactions, we focus here on short-range
interactions with $J\left( \bf{k}\right) =J\sum_{\mathbf{\delta}}e^{i\bf{k}\cdot \mathbf{\delta }}$ and hence $%
\gamma _{\bf{k}}=\frac{1}{z}\sum_{\bf{ \delta }}e^{i%
\bf{k}\cdot \mathbf{\delta }}$ where $z$ is the
coordination number and $\mathbf{\delta}$ denotes the nearest-neighbor spin vectors.

In such a case, it is simple to show that \cite{callen1963}
\begin{equation}
\sum_{\bf{k}^{^{\prime }}}\left( \gamma _{\bf{k}%
^{^{\prime }}}-\gamma _{\bf{k}-\bf{k}^{^{\prime
}}}\right) \varphi \left( \bf{k}^{^{\prime }}\right) =\left(
1-\gamma _{\bf{k}}\right) \sum_{\bf{k}^{^{\prime
}}}\gamma _{\bf{k}^{^{\prime }}}\varphi \left( \bf{k}%
^{^{\prime }}\right) .  \label{sc5}
\end{equation}
Then, the basic equations become
\begin{equation}
\left\{
\begin{array}{ccc}
\widetilde{\omega }_{\bf{k}} &=& \widetilde{h}+\sigma \left(
1-\gamma _{\bf{k}}\right) R \\
\sigma  &=& L\left( \frac{1}{\Phi}\right)
\end{array}
\right. ,  \label{sc6}
\end{equation}
with
\begin{equation}
\Phi =\widetilde{T}\int_{1BZ}\frac{d^{d}k}{\left( 2\pi \right)
^{d}}\frac{1}{\widetilde{h}+\sigma \left( 1-\gamma _{\bf{k}%
}\right) R},  \label{sc7}
\end{equation}
where now $R$ is independent of $\bf{k}$ and given by the
equation
\begin{equation}
R=1+\lambda \sigma \widetilde{T}\int_{1BZ}\frac{d^{d}k}{\left( 2\pi \right)
^{d}}\frac{\gamma _{\bf{k}}}{\widetilde{h}+\sigma \left(
1-\gamma _{\bf{k}}\right) R}.  \label{sc8}
\end{equation}
Of course, this quantity depends on the decoupling used to close the
equation of motion for the original GF ($\lambda =0$ for TD and $\lambda =1$
for CD).

In the limit $\widetilde{h}\rightarrow 0^{+}$ with $\sigma \geq 0$, Eqs. (%
\ref{sc6})-(\ref{sc7}) reduce to
\begin{equation}
\left\{
\begin{array}{ccc}
\widetilde{\omega }_{\bf{k}}&=&\sigma \left(
1-\gamma _{\bf{k}}\right) R \\
\sigma &=&\coth\left( \frac{\sigma }{\widetilde{T}Q\left( \widetilde{T}\right) }%
\right) -\left( \frac{\widetilde{T}Q\left( \widetilde{T}\right) }{\sigma }%
\right)
\end{array}
\right. ,  \label{sc9}
\end{equation}
with
\begin{eqnarray}
\Phi &=&\frac{\widetilde{T}Q\left( \widetilde{T}\right) }{\sigma },
\\ \label{sc10}
Q\left( \widetilde{T}\right) &=&\frac{F_{d}\left( -1\right) }{R},  \label{sc11}
\end{eqnarray}
and
\begin{equation}
R=1+\lambda \frac{\widetilde{T}}{R}\widetilde{F}_{d}\left( -1\right) .
\label{sc12}
\end{equation}
Here
\begin{eqnarray}
F_{d}\left( -1\right) &=&\int_{1BZ}\frac{d^{d}k}{\left( 2\pi \right) ^{d}}%
\frac{1}{\left( 1-\gamma _{\bf{k}}\right) }, \label{sc13a}\\
\widetilde{F}_{d}\left( -1\right) &=&\int_{1BZ}\frac{d^{d}k}{\left( 2\pi
\right) ^{d}}\frac{\gamma _{\bf{k}}}{\left( 1-\gamma _{%
\bf{k}}\right) }=F_{d}\left( -1\right) -1,\label{sc13b}
\end{eqnarray}
from the conventional notation for the so-called ``structure sums'' $%
F_{d}\left( n\right) =\frac{1}{N}\sum_{\bf{k}}\left( 1-\gamma _{%
\bf{k}}\right) ^{n}$ depending only on the lattice structure of
the spin model.

From Eq. (\ref{sc12}) one immediately obtains the physical solution for $R$ (%
$\geq 1$):
\begin{equation}
R=\frac{1}{2}\left[ 1+\sqrt{1+4\lambda \widetilde{T}\widetilde{F}_{d}\left(
-1\right) }\right] .  \label{sc14}
\end{equation}
With this expression, the basic equations for the zero-magnetic field
problem assume the form
\begin{equation}
\left\{
\begin{array}{c}
\widetilde{\omega }_{\bf{k}}\left( \widetilde{T}\right) =\frac{1%
}{2}\sigma \left( \widetilde{T}\right) \left[ 1+\sqrt{1+4\lambda \widetilde{T%
}\widetilde{F}_{d}\left( -1\right) }\right] \left( 1-\gamma _{%
\bf{k}}\right)  \\
\sigma \left( \widetilde{T}\right) =\coth\left( \frac{\sigma \left( \widetilde{%
T}\right) }{\widetilde{T}Q\left( \widetilde{T}\right) }\right) -\left( \frac{%
\widetilde{T}Q\left( \widetilde{T}\right) }{\sigma \left( \widetilde{T}%
\right) }\right)
\end{array}
\right. ,  \label{sc15}
\end{equation}
where
\begin{equation}
Q\left( \widetilde{T}\right) =\frac{2F_{d}\left( -1\right) }{\left[ 1+\sqrt{%
1+4\lambda \widetilde{T}\widetilde{F}_{d}\left( -1\right) }\right] }.
\label{sc16}
\end{equation}
Notice that for $\lambda =0$ (TD) the quantity $Q$ does not depend on $%
\widetilde{T}$. 

Previous equations show that the zero-magnetic field problem
reduces to solve the single self-consistent equation (\ref{sc9}) for $\sigma
\left( \widetilde{T}\right) $ for dimensionalities for which the integrals (%
\ref{sc13a}) and (\ref{sc13b}) converge and hence long-range order may occur.

Now, we have all the ingredients to explore the relevant thermodynamic
properties of our classical model for different values of dimensionality.
This will be the subject of the next section.

It is worth emphasizing that previous equations differ from those obtained
in Ref. \cite{prof1984} only for the expression of magnetization. The use of this formula avoids making the assumptions which were not sufficiently justified and tested almost
thirty years ago in the context of the CSDM.

\section{Magnetization calculations}~\label{calculations}

By inspection of Eqs. (\ref{sc2})-(\ref{sc4}) or (\ref{sc6})-(\ref{sc8}),
we see that at zero temperature it is $\Phi=0$ for any $d$ and
hence $\sigma =1$  for any value of $\widetilde{h}\geq 0$ (fully
polarized ground state). On the other hand, for finite temperature and $%
\widetilde{h}=0$, since $1-\gamma _{\bf{k}}\simeq \frac{1}{z}%
k^{2}$ as $k\rightarrow 0$ ($J\left( 0\right) -J\left( \bf{k}%
\right) \simeq \frac{1}{z}J\left( 0\right) k^{2}$ as $k\rightarrow 0$), the
integrals (\ref{sc13a}) and (\ref{sc13b}) converge only for $d>2$ where
long-range order (LRO) is expected, consistently with the classical version
\cite{Bruno} of the Mermin-Wagner theorem \cite{Mermin1}. For $d\leq 2$ one
has $\Phi=\infty $ at $\widetilde{h}=0$ and no LRO may occur at
finite temperature, i. e. $\sigma \left( \widetilde{T}\right) \equiv 0$,
defining a paramagnetic phase. In this asymptotic scenario, $d=2$
assumes the role of lower critical dimension for the classical Heisenberg
model.

In the low-temperature regime, close to the polarized state, we have $%
\Phi\ll 1$ ($\Phi\rightarrow 0$ as $\widetilde{T}%
\rightarrow 0$) so that $\coth\left( \frac{1}{\Phi }\right) \simeq
1+O\left( e^{-2/\Phi }\right) $ and Eq. (\ref{sc6}) provides $\sigma
\simeq 1-\Phi +O\left( e^{-2/\Phi }\right) $. So,
Eqs. (\ref{sc6})-(\ref{sc8}) reduce exactly to those explored in detail in Refs. \cite
{prof1984,cavallo,prof1,rassegna2006} and here we do not consider again this regime.

Besides, at high temperature we have $\Phi \gg 1$, which implies
$\sigma \ll 1$, and the asymptotic results derived in Refs. \cite
{prof1984,rassegna2006} are reproduced.

Then, the properties for intermediate values of $\Phi$ are of
main interest for us since, in this crossover regime, the expression for
$\sigma $ obtained in the present paper (see Eq. (\ref{sc2})) differs
analytically from the corresponding one (in our notations) $%
\sigma =\left\{ \frac{1-3\sigma \Phi }{1-\sigma \Phi %
}\right\} ^{1/2}$ suggested in Ref. \cite{prof1984} within the framework of
the CSDM. This was successfully employed  in the recent study \cite{prb2010}
about more complex classical Heisenberg models obtaining accurate
results in surprising agreement with experiments.

In the following part of this section, we limit ourselves to consider the
cases $d=1$ and $d=3$ for which suitable results exist~\cite{prof1984,tognetti1} for a meaningful comparison with our predictions
based on Eq. (\ref{sc2}).

\subsection{The classical ferromagnetic chain}

As mentioned before, the one-dimensional model does not exhibit LRO ($\sigma
\left( \overline{T}\right) \equiv 0$ in zero magnetic field) and the
quantity of theoretical and experimental interest is the paramagnetic
susceptibility $\widetilde{\chi }=\lim_{\widetilde{h}\rightarrow 0}\frac{%
\sigma }{\widetilde{h}}$.

In the asymptotic regimes $\widetilde{T}\gg 1$ and $\widetilde{T}\ll 1$ with
$\Phi\gg 1$ we find, respectively,
\begin{equation}
\widetilde{\chi }\simeq \left\{
\begin{array}{cc}
\frac{1}{3\widetilde{T}}\left\{ 1+\frac{1}{3\widetilde{T}}+\frac{2}{3}\left[
\frac{1}{3\widetilde{T}}\right] ^{2}+...\right\} , & \widetilde{T}\gg 1, \\
\frac{8}{3}\left[ \frac{1}{3\widetilde{T}}\right] ^{2}, & \widetilde{T}\ll 1,
\end{array}
\right.   \label{ferr1}
\end{equation}
which coincide, as expected, with the corresponding ones derived 
in Ref. \cite{prof1984} in the same regime ($\Phi\gg 1$).
Remarkably, these results differ from the exact ones \cite{fisher1} with $%
\frac{1}{2}$ and $3$ instead of $\frac{2}{3}$ and $\frac{8}{3}$ in Eq. (\ref
{ferr1}).

Similarly, in the regime $\Phi \ll 1$, near the polarized state,
we find the same results reported in Ref. \cite{prof1984}.

As already noted at the beginning of this section, only when $\widetilde{h}%
\neq 0$ and $\Phi$ is finite, one can expect to find
differences from our framework and that of Ref. \cite{prof1984} at the same
level of decoupling approximation.

\begin{figure}[th]
\centering\includegraphics*[width=0.8\linewidth]{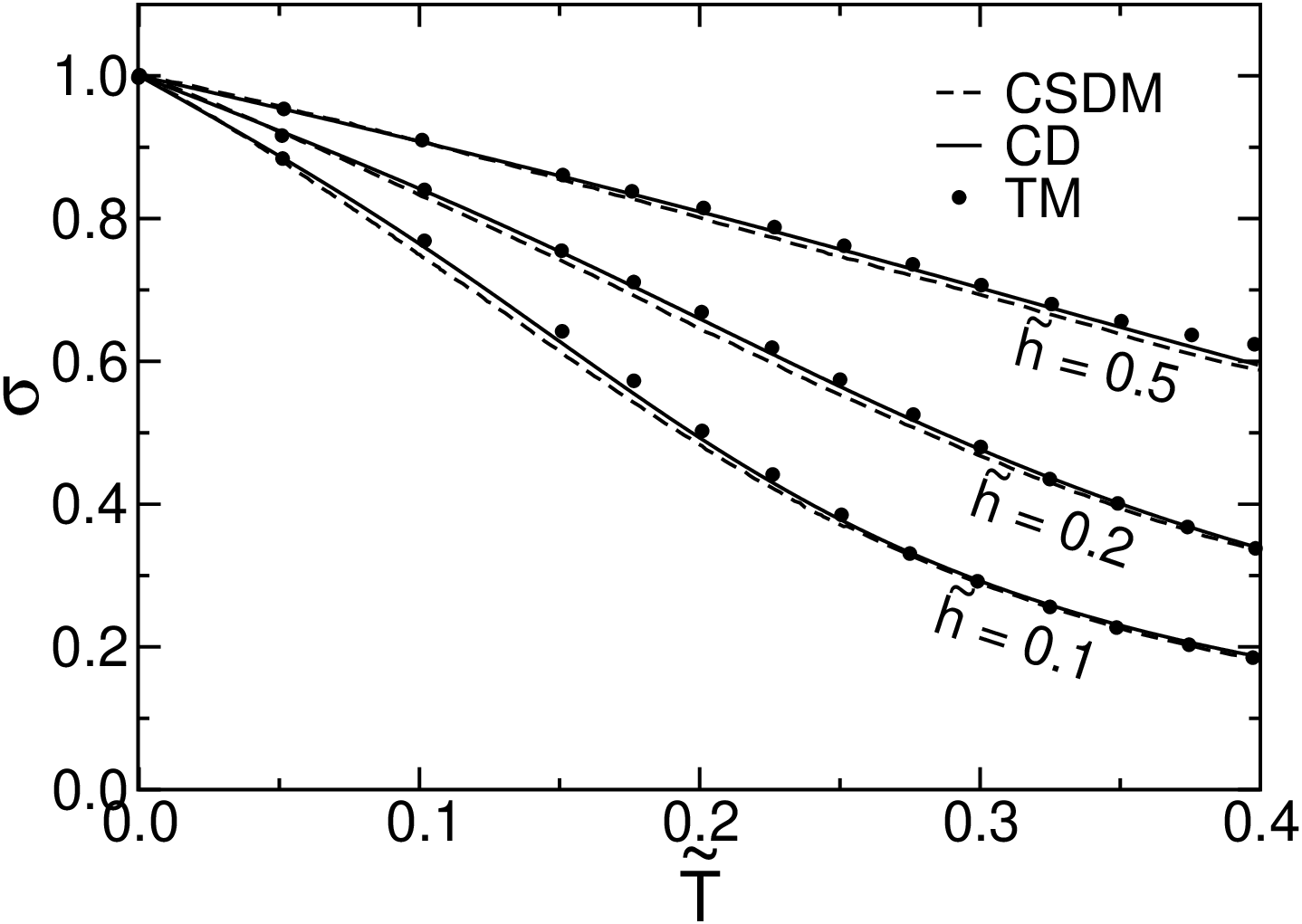}
\caption{Plots of  magnetization $\protect\sigma $ as a function of $%
\widetilde{T}$ for different values of the reduced magnetic field $%
\widetilde{h}$. The solid and dashed lines represent the Callen-like
decoupling (CD) results and those derived in Ref. \protect\cite{prof1984}, respectively. The dots denote the exact transfer-matrix data.}
\label{figura1}
\end{figure}

Unfortunately, a comparison of corresponding results can be performed only
numerically. The reduced magnetization $\sigma $, derived by Eqs. (\ref{sc6}%
)-(\ref{sc8}) as a function of $\widetilde{T}$ and $\widetilde{h}$ within
the CD framework is plotted in Fig.~\ref{figura1} and compared with the
corresponding predictions obtained by means of CSDM with $\sigma =\left\{
\frac{1-3\sigma \Phi}{1-\sigma \Phi}\right\} ^{1/2}
$ and with the exact numerical transfer-matrix data \cite{tognetti2}.

Remarkably, at finite temperature and higher magnetic fields, our CD results
are nearer to the exact numerical transfer-matrix data than those found with
the CSDM \cite{prof1984}, although they are very near and essentially
coincident at near zero and high temperatures.

A further signature of the greatest accuracy of the present results at finite
values of $\Phi$ arises from numerical calculations for the
transverse correlation length defined by \cite{prof1984,tognetti2}:
\begin{equation}
\xi _{\perp }=-\frac{1}{2}\lim_{k\rightarrow 0}\left[ \left(\frac{d^{2}}{dk^{2}}%
\left\langle S_{\bf{k}}^{+}S_{-\bf{k}%
}^{-}\right\rangle\right) \diagup \left\langle S_{\bf{k}}^{+}S_{-%
\bf{k}}^{-}\right\rangle \right] .  \label{ferr2}
\end{equation}
In general, using the expression of $\left\langle S_{\bf{k}}^{+}S_{-\bf{k}%
}^{-}\right\rangle $ in terms of $\sigma $, one easily finds within our
approximation and in $d$ dimensions
\begin{equation}
\xi _{\perp }=\left( \frac{\sigma R}{z\widetilde{h}}\right) ^{\frac{1}{2}},
\label{ferr3}
\end{equation}
which should provide information also about the paramagnetic susceptibility $%
\widetilde{\chi }=\frac{\sigma }{\widetilde{h}}$ decreasing to zero the
magnetic field.

Numerical CD results of Eq. (\ref{ferr3}) for $d=1$ with $z=2$ are plotted
in Fig. 2 and compared with other data as in Fig.~\ref{figura1}.

\begin{figure}[th]
\centering\includegraphics*[width=0.8\linewidth]{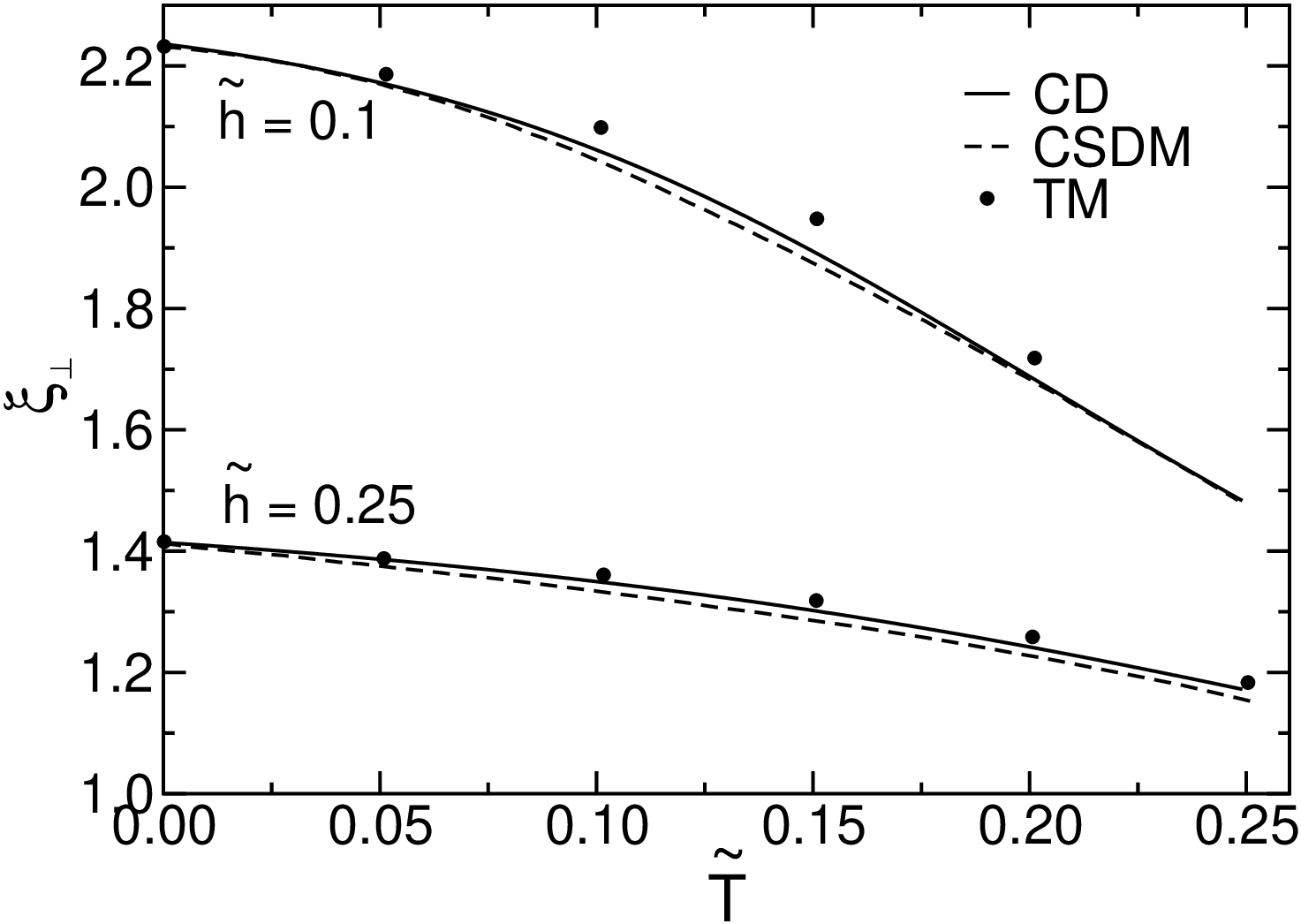}
\caption{Plot of the transverse correlation length $\protect\xi _{\bot }$ as
a function of reduced temperature $\widetilde{T}$ for different values of
the reduced magnetic field $\widetilde{h}$. The solid and dashed lines stand
for the present CD-predictions and the CSDM ones of Ref.
\protect\cite{prof1984}, respectively. The dots denote the exact
transfer-matrix data.}
\label{figura2}
\end{figure}

Notice that our results appear again more accurate at intermediate values of
$\Phi $ reproducing more effectively the exact numerical transfer-matrix
results.

\subsection{$d>2$ and three-dimensional ferromagnetic models}

For all $d>2$, LRO is possible and the zero-magnetic field equations (\ref
{sc14})-(\ref{sc16}) allow to find the reduced critical temperature $%
\widetilde{T}_{c}$ and the spontaneous magnetization $\sigma \left(
\widetilde{T}\right) $ in the full range $0<\widetilde{T}<\widetilde{T}_{c}$.

First, we note that, for $\sigma \rightarrow 0^{+}$ (with $\widetilde{T}%
\rightarrow \widetilde{T}_{c}^{-}$ and $\widetilde{T}_{c}$ to be determined),
we have $\Phi=\frac{\widetilde{T}Q\left( \widetilde{T}\right) }{%
\sigma \left( \widetilde{T}\right) }\rightarrow \infty $. Thus, using the
expansion $\coth x-\frac{1}{x}=\sum_{n=1}^{\infty }\frac{2^{2n}B_{2n}}{\left(
2n\right) !}x^{2n-1}$ where $B_{2n}$ are the Bernoulli numbers, we can write
for $\sigma \ll 1$ the equation
\begin{equation}
\sigma \simeq \frac{1}{3}\left( \frac{\sigma }{\widetilde{T}Q\left(
\widetilde{T}\right) }\right) -\frac{1}{45}\left( \frac{\sigma }{\widetilde{T%
}Q\left( \widetilde{T}\right) }\right) ^{3}+...\quad.  \label{ferr4}
\end{equation}
Then, we get
\begin{equation}
\begin{array}{cc}
\sigma \left( \widetilde{T}\right) \simeq \sqrt{15}\left( \widetilde{T}%
Q\left( \widetilde{T}\right) \right) \left( 1-3\widetilde{T}Q\left(
\widetilde{T}\right) \right)^\frac{1}{2} , & \widetilde{T}\rightarrow \widetilde{T}_{c}
\end{array}
\label{ferr5}
\end{equation}
and $\widetilde{T}_{c}$ is determined by the equation
\begin{equation}
1-3\widetilde{T}_{c}Q\left( \widetilde{T}_{c}\right) =0.  \label{ferr6}
\end{equation}
So, from Eq. (\ref{ferr5}), we have
\begin{equation}
\sigma \left( \widetilde{T}\right) \simeq \sqrt{\frac{5}{3}}\left( 1+%
\widetilde{T}_{c}\frac{Q^{^{\prime }}\left( \widetilde{T}_{c}\right) }{%
Q\left( \widetilde{T}_{c}\right) }\right) \left( \frac{\widetilde{T}_{c}-%
\widetilde{T}}{\widetilde{T}_{c}}\right) ^{1/2},  \label{ferr7}
\end{equation}
which defines the mean-field approximation (MFA) critical exponent $\beta =%
\frac{1}{2}$.

The explicit expression of $\widetilde{T}_{c}$ for $d>2$ can be immediately
determined form Eqs. (\ref{sc16}) and (\ref{ferr6}) providing
\begin{equation}
\widetilde{T}_{c}=\frac{1}{3F_{d}\left( -1\right) }\left\{ 1+\frac{\lambda }{%
3}\frac{\widetilde{F}_{d}\left( -1\right) }{F_{d}\left( -1\right) }\right\} .
\label{ferr8}
\end{equation}
In particular, we get
\begin{equation}
\widetilde{T}_{c}^{\left( TD\right) }=\frac{1}{3F_{d}\left( -1\right) },
\label{ferr9}
\end{equation}
\begin{equation}
\widetilde{T}_{c}^{\left( CD\right) }=\widetilde{T}_{c}^{\left( TD\right)
}\left\{ 1+\frac{1}{3}\frac{\widetilde{F}_{d}\left( -1\right) }{F_{d}\left(
-1\right) }\right\} >\widetilde{T}_{c}^{\left( TD\right) },  \label{ferr10}
\end{equation}
similarly to the Callen result for the quantum Heisenberg model \cite{callen1963}.

The numerical values of $\widetilde{T}_{c}$ for the three-dimensional Heisenberg ferromagnet, as obtained by means of
different methods, are
reported in Table I for simple cubic (sc) - body centered cubic (bcc) - and
face centered cubic (fcc) spin lattices.
\begin{table}
\caption{ Numerical values of the reduced critical temperature $\widetilde{T}%
_{c}$ of the three-dimensional classical Heisenberg model for different
lattice structures. HTS stands for the exact high-temperature-series results.}
\begin{center}
\begin{tabular*}{0.62\textwidth}{c|c|c|c|c|c}
\hline
$\widetilde{T}_{c}$ & MFA  & CD & TD & CSDM & HTS \\
\hline
\hline
sc($z=6$) & 0.333& 0.245 & 0.220 & 0.245 & 0.241\\
bcc($z=8$)& 0.333& 0.262 & 0.239 & 0.262 & 0.257\\
fcc($z=12$)& 0.333 & 0.269 & 0.248 & 0.269 & 0.265\\
\hline
\end{tabular*}
\end{center}
\end{table}

Notice that our CD-results, as the identical CSDM ones, are very close to
the exact HTS results of Ref.~\cite{rush1}.

As concerning the spontaneous magnetization $\sigma \left( \widetilde{T}%
\right) $ as a solution of the self-consistent equation (\ref{sc15}), one expects that
the CD results will differ from the CSDM~\cite{prof1984} ones in a temperature range
sufficiently far from the asymptotic regimes near zero ($\Phi %
\gg 1$) and critical ($\Phi \ll 1$) temperatures.

Eq. (\ref{sc15}) can be solved only numerically for any temperature in the
interval $0<\widetilde{T}<\widetilde{T}_{c}$. The data for $\sigma \left(
\widetilde{T}\right) $ of the $\left( d=3\right) $-classical Heisenberg
ferromagnet for different lattice structures are plotted in Fig. 3 and
compared with the CSDM and MFA results.

\begin{figure}[bh]
\centering\includegraphics*[width=0.32\linewidth]{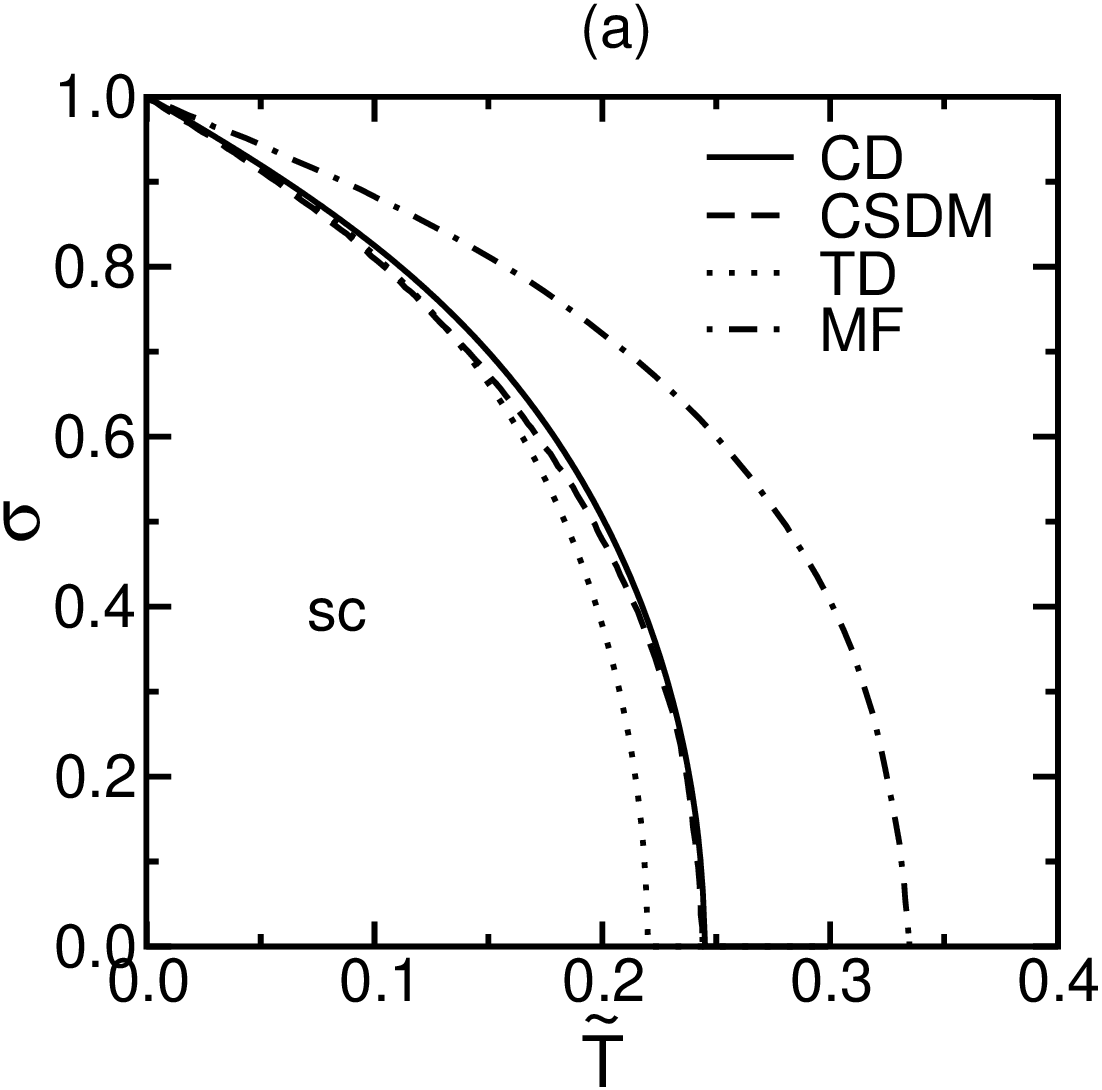}
\centering\includegraphics*[width=0.32\linewidth]{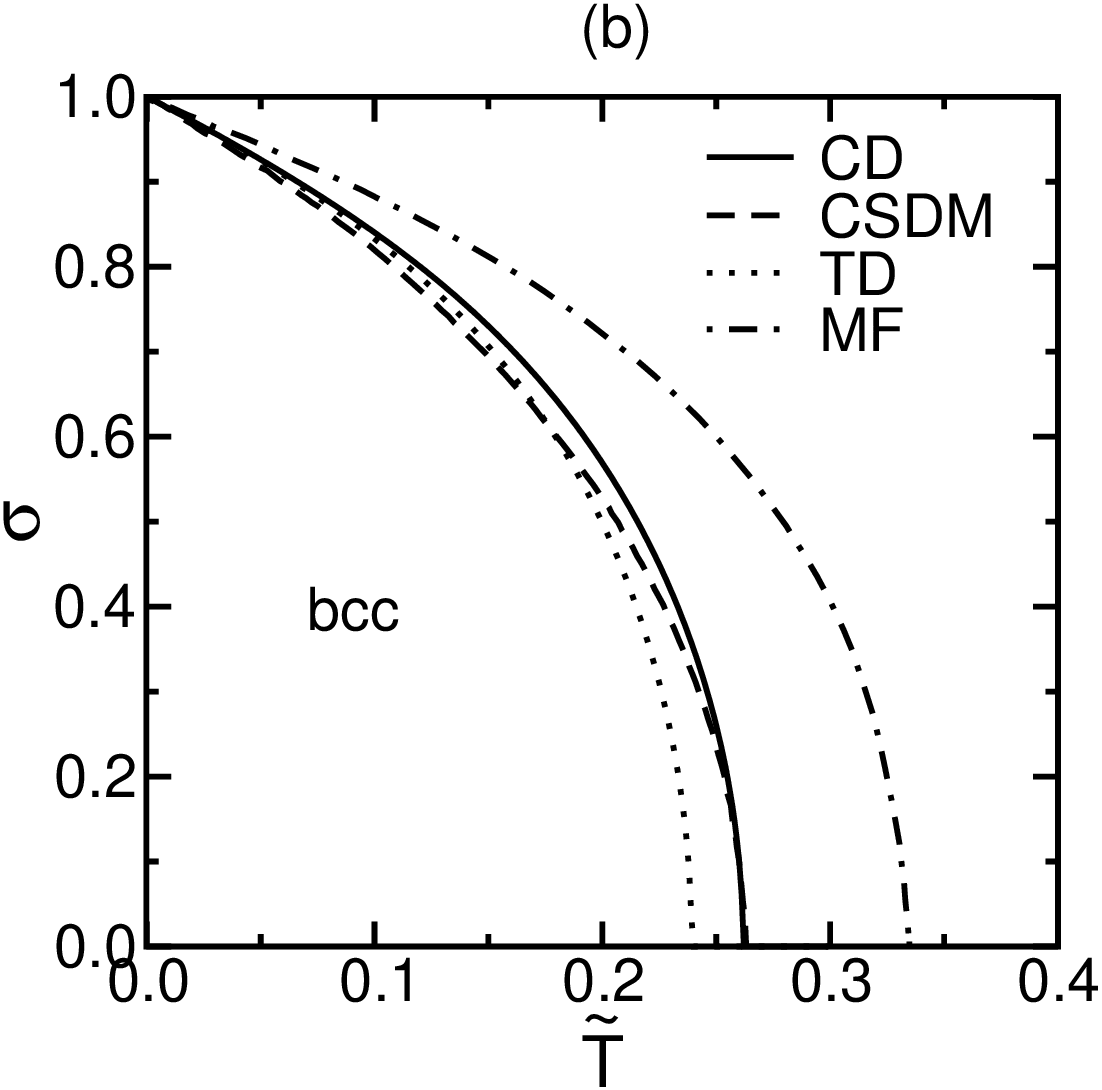}
\centering\includegraphics*[width=0.32\linewidth]{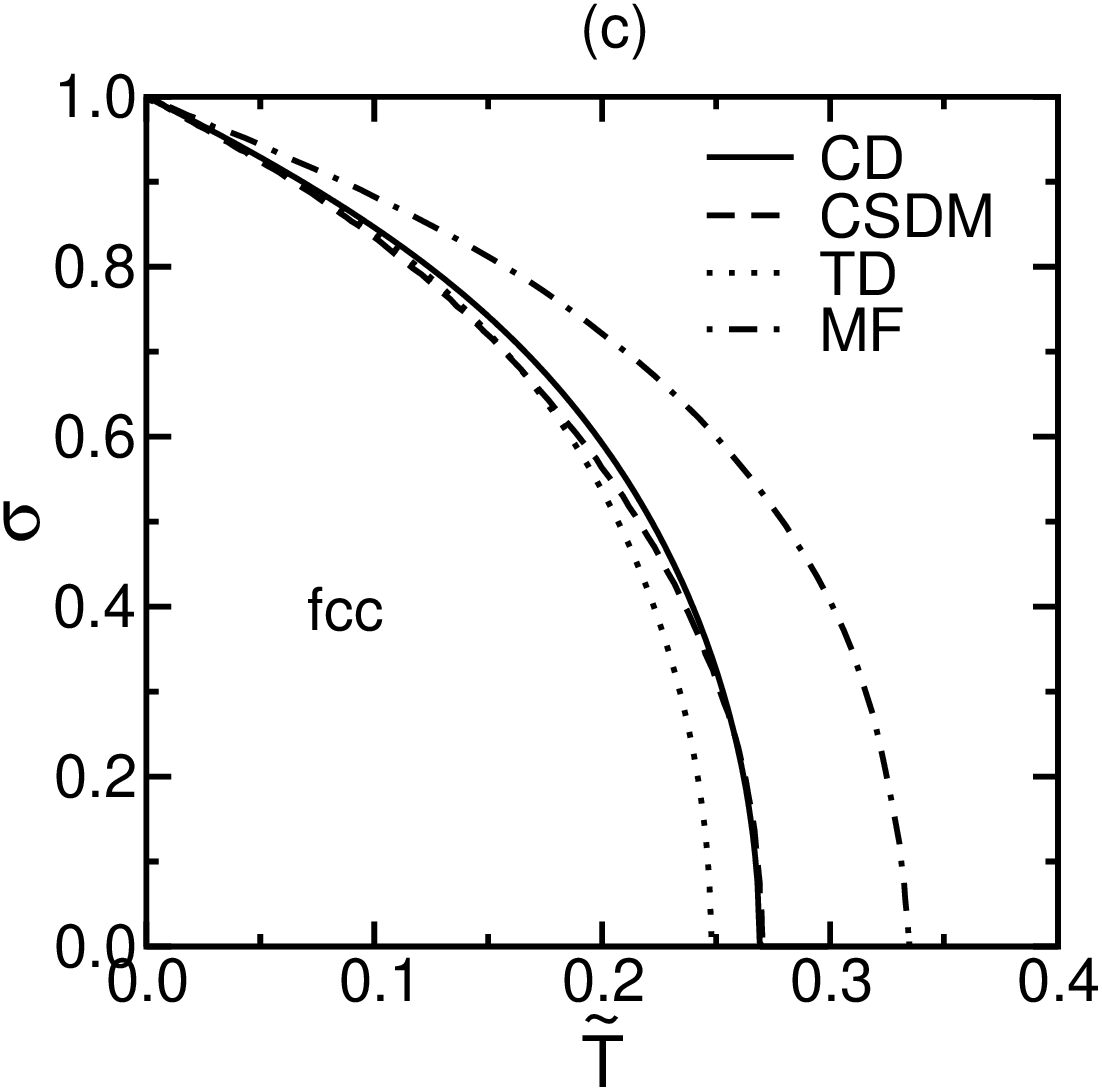}
\caption{Plots of magnetization $\protect\sigma $ as a function
of the reduced temperature $\widetilde{T}$ for different three-dimensional
lattices: (a) simple cubic (sc); (b) body centered cubic (bcc); (c) face
centered cubic (fcc). Comparison is made between present TD and CD
calculations and the CSDM and MFA corresponding ones.}
\label{figura3}
\end{figure}

It is worth emphasizing again that, as expected, differences between our
results and the CSDM ones occur for finite temperature in the range $%
0.05\lesssim \widetilde{T}\lesssim 0.25$.

\section{Conclusions}~\label{conclusions}

The main purpose of the present paper was the derivation of a new formula for the magnetization of the $d$-dimensional classical Heisenberg ferromagnetic model. This is a central problem within the two-time GF framework and the strictly related spectral density method in classical statistical physics. The main difficulty indeed is crucially due to the absence of the classical counterpart of the exact kinematic rule for the $z$-components of the quantum spin vectors.
By a suitable extension of the Callen method developed almost five decades ago for a spin-S quantum Heisenberg ferromagnet~\cite{callen1963}, this new formula has been obtained analytically overcoming the previously mentioned difficulty. Surprisingly, although it is formally different from that suggested within the CSDM, both the frameworks provide results in good agreement with the exact numerical transfer-matrix data available for the classical Heisenberg ferromagnetic chain and with the exact HTS ones for three- dimensional case.
 However, the present formula provides more accurate results in an appreciable range of temperatures and magnetic field where the differences between the two formulas become evident and suitably interpolates the low- and high- temperature regimes. This emerges from a careful comparison between our predictions and those numerically accessible for the magnetization and other directly related static quantities as the transverse correlation length and the paramagnetic susceptibility. In any case both formulas work surprisingly well.
Of course, other experimentally relevant thermodynamic quantities, such as the internal and free energies and the specific heat at fixed magnetic field, can be found by using  the general classical two-time GF formalism introduced by Bogoliubov Jr and Sadovnikov~\cite{bogoliubov} and further developed in successive papers~\cite{prof1984,cavallo,prof1,rassegna2006}, in strict analogy with the better known quantum counterpart, but these calculations are beyond the purposes of the present paper.

In conclusion, we believe that our new formula for magnetization may be usefully employed for further theoretical and numerical calculations to explore the properties of other Heisenberg-type classical spin systems at different dimensionalities.

\section*{Acknowledgments}
A.C. acknowledges the MIUR (Italian Ministry of Research) for financial support
within the program ``Incentivazione alla mobilit\`a di studiosi stranieri e italiani residenti all'estero''.

\end{document}